\documentclass[referee]{raa}            % referee version: for submission

\usepackage{graphicx,times}             %for PS/EPS graphics inclusion, new
\usepackage{natbib}
\usepackage{amssymb,amsmath}
\bibpunct{(}{)}{;}{a}{}{,}

\usepackage[]{hyperref}

\begin{document}

   \title{Partitioning the Galactic Halo with Gaussian Mixture Models
}
%   \subtitle{I. Place Your Subtitle Here}

   \volnopage{Vol.0 (20xx) No.0, 000--000}      %%preserved for Editor. DOn't remove!
   \setcounter{page}{1}          %%starting page, preserved for Editor. DOn't remove!

   \author{Xilong Liang
      \inst{1,2}
   \and Yuqin Chen
      \inst{2,1}
   \and Jingkun Zhao
      \inst{2,1}
   \and Gang Zhao
      \inst{2,1}
   }

   \institute{School of Astronomy and Space Science, University of Chinese Academy of Sciences, Beijing 100049, China\\
        \and
             CAS Key Laboratory of Optical Astronomy, National Astronomical Observatories, Chinese Academy of Sciences,Beijing 100101, China; {\it cyq@bao.ac.cn}\\
\vs\no
   {\small Received~~20xx month day; accepted~~20xx~~month day}}
\abstract{ The Galactic halo is supposed to form from merging with nearby dwarf galaxies. In order to probe different components of the Galactic halo, we have applied the  Gaussian Mixture Models method to a selected sample of metal poor stars with [Fe/H] $< -0.7$ dex in the APOGEE DR16 catalogue based on four-parameters, metallicity, [Mg/Fe] ratio and spatial velocity (\textit{$V_R$, $V_\phi$}). Nine groups are identified with four from the halo (group 1, 3, 4 and 5), one from the thick disk (group 6), one from the thin disk (group 8) and one from dwarf galaxies (group 7) by analyzing their distributions in the ([M/H], [Mg/Fe]), ($V_R$, $V_\phi$), (\textit{Zmax}, \textit{eccentricity}), (\textit{Energy}, \textit{Lz}) and ([Mg/Mn], [Al/Fe]) coordinates. The rest two groups are respectively caused by observational effect (group 9) and the cross section component (group 2) between the thin disk and the thick disk. It is found that in the extremely outer accreted halo (group 1), stars born in the Milky Way can not be distinguished from those accreted from other galaxies either chemically or kinematically. In the intermediate metallicity of $-$1.6 $<$ [Fe/H] $< -0.7$ dex, the accreted halo mainly composed of the Gaia-Enceladus-Sausage substructure (group 5), which can be easily distinguished from group 4 (the in-situ halo group) in both chemical and kinematic space. Some stars of group 4 may come from the disk and some disk stars can be scattered to high orbits by resonant effects as shown in the \textit{Zmax} versus Energy coordinate. We also displayed the spatial distribution of main components of the halo and the ratio of accreted components do not show clear relation to Galactic radius.
}

   \authorrunning{Liang et al. }            %author_head in even pages
   \titlerunning{Halo groups }  % title_head in odd pages

   \maketitle
%% The author head (on even pages) and the title head (on odd pages) will be
%% automatically extracted from \author{} and \title{}. Whenever the title is too long,
%% you will be asked to supply a shorter one by inserting either \authorrunning{} or
%% \titlerunning{} before \maketitle. Anyway, you can specify your own heads.
%%
%%
%% Note: In the following text body of your manuscript, please note several differences from
%%       other major journals:
%% (1) \subsection{Please Capitalize the First Letter of Each Notional Word in Subsection Title}
%% (2) Please Capitalize the First Letter of Each Notional Word in all tables' captions

%
%________________________________________________ sections below
%
\section{Introduction}
According to the $\Lambda$ cold dark matter scenario, the halos of large galaxies like the Milky Way grow in size by merging with small dwarf galaxies, leaving behind debris in the form of different groups of stars \citep{fre12,2020arXiv200204340H}.  With Gaia second data release \citep{gaia18} and large spectroscopic surveys (LAMOST \citep{2012RAA....12.1197C,2012RAA....12..723Z}; RAVE \citep{2017AJ....153...75K}; APOGEE \citep{2017AJ....154...94M}; GALAH \citep{2018MNRAS.478.4513B} and so on), the Galactic halo has been analysed both kinematically and chemically. A number of studies have drawn a consistent picture that a large fraction of stellar halo was accreted. One recently found velocity structure has been named Gaia-Enceladus-Sausage \citep{bel18,dea18,hay18,hel18,kop18,mye18c,mye18d,fat19}, which is speculated to be debris from a massive dwarf galaxy (Gaia Enceladus) with initial stellar mass about $ 5 \times 10^{8} - 5 \times 10^{9} M_{\bigodot} $ \citep{bel18,hel18,mac19,vin19} and it was supposed to be accreted about 10 Gyr ago \citep{hel18,dim18,gal19}. Besides Gaia-Enceladus-Sausage, \citet{mye18c} identified another velocity structure consisting of retrograde, high-energy stars in the halo with metallicity between $-1.9$ and $-1.3$, which is related to $\omega$ Centauri by \cite{mye18b,mye18c} and is supposed to be accreted $5-8$ Gyr ago \citep{kop19}. According to \cite{bek03,maj12}, dwarf galaxy $\omega$ Centauri is known to be a major source of retrograde halo stars in the inner Galaxy. \citet{mas19} argues that Gaia-Enceladus-Sausage is also likely associated with $\omega$ Centauri. A new high-energy, retrograde velocity structure named as term Sequoia is found by \citet{mye19}. It is connected with a large globular cluster with very retrograde halo-like motion, FSR-1758 \citep{bar19}. Based on chemical abundances, \citet{mat19} suggested that this retrograde component is dominated by an accreted dwarf galaxy, which has a longer star formation timescale and is less massive than Gaia-Enceladus-Sausage. In summary, \citet{kop19} suggested that the retrograde halo contains a mixture of debris from objects like Gaia-Enceladus-Sausage, Sequoia, Thamnos and even the chemically defined thick disc. Based on chemical abundances, \citet{nis10} divided the halo into high $\alpha$ population and low $\alpha$ population and they proposed that the low $\alpha$ population is accreted from nearby galaxies. Based on SDSS data, \citet{2007Natur.450.1020C,car10} classified the halo into the inner halo with metal rich stars on mildly eccentric orbits and the outer accreted halo with metal poor stars on more eccentric orbits. From APOGEE DR12, \citet{haye18} found that the low [Mg/Fe] population has a large velocity dispersion with very little or no net rotation. \citet{hay18} suggested that the high $\alpha$ stars thought to belong to an in situ formed halo population may in fact be the low rotational velocity tail of the old Galactic disk heated by the last significant merger of a dwarf galaxy. \citet{dim19} also claimed that about half of the kinematically defined halo within a few kilo parsec from the Sun is composed of thick disc stars, since the accretion process could lead to the last significant heating of the thick disk stars into the halo. With [M/H], [Mg/Fe] and distances from APOGEE data release 14, \citet{chen19} indicates a three section halo, the inner in situ halo with $|Z|$ less than about $8 - 10$ kpc, the intermediately outer dual-mode halo at $|Z|$ between about 10 kpc and 30 kpc, and the extremely outer accreted halo with $|Z|$ larger than 30 kpc. \citet{2020arXiv201000235C} found the inner stellar halo is also comprised of many different components. In a word, the Galactic halo has a complicate history and further works are desired to unravel its assembling processes.

This paper aims to partition the Galactic halo by identifying different components in the chemical and kinematic space. As shown in \citet{nis10} and \citet{hay18}, in the intermediate metallicity range of $-1.6$ $<$ [Fe/H] $< -0.7$ dex,  the accreted halo can be separated from the in-situ halo in the [M/H] versus [Mg/Fe] coordinate because there is a clear gap between low [Mg/Fe] and high [Mg/Fe] stars. Those two populations also have distinguished kinematic properties in the Toomre diagram \citep{nis10}, the $V_R$-$V_\phi$  diagram \citep{hay18} and eccentricity distributions \citep{mac19}. Note that, the separation between the accreted halo and the in-situ halo is not found in the chemical space of [M/H] versus [Mg/Fe] in the low metallicity range of $-2.6$ $<$ [Fe/H] $<-1.6$ dex in the APOGEE data. It would be of interest to investigate whether the kinematic imprints of the accreted halo can be picked out for this low metallicity range by using a grouping method that simultaneously takes into account chemical and kinematic properties. Finally, after groups are selected, studying similarities and differences among different groups by various combinations of orbital and energy parameters provide new insights on the complicate assembling history of the Galaxy.

\section{Data and the partition method}

The sample stars are selected from the APOGEE DR16 dataset \citep{abo18} with parallaxes and proper motions taken from Gaia DR2 \citep{gaia18}. We have removed stars with ASCAPFLAG or STARFLAG warnings and stars with $-9999$ values of $\log g$, $T_{\textrm{eff}}$, [Fe/H] and [Mg/Fe]. We chose to use the Bailer-Jones distance \textit{GAIA\_R\_EST} \citep{bai18} and removed stars with $\frac{(GAIA\_R\_HI - GAIA\_R\_LO)}{2}/GAIA\_R\_EST > 0.2$. After these cuts, there are 165 332 stars left. Then, we calculated three dimensional velocity components in galactocentric cylindrical coordinate with radial velocities from APOGEE DR16. Python package astropy has been used to transform observed quantities in ICRS coordinate into galactocentric cylindrical coordinate. The Sun is placed at height $z = 0.014$ kpc, galactic radius $R = 8.2$ kpc with circular speed $ V_{c} = 233.1$ km\,s$^{-1}$ \citep{mcm11}. The peculiar velocity of the Sun relative to the local standard of rest is taken as $(U_{\bigodot}, V_{\bigodot}, W_{\bigodot}) = $ (11.1, 12.24, 7.25) km\,s$^{-1}$ from \citet{sch10}. Since we want to analyse main components of the halo in the field, we removed those stars with PROGRAMNAMEs in the APOGEE DR16 catalogue associated with globular clusters, bulge, young stellar object, RR Lyrae stars, exoplanets, the Magellanic cloud or open clusters. Moreover, stars with PROGRAMNAME related to stellar streams and apparently clumped as a small group in the velocity coordinate have been removed. After those observationally clumped stars removed, there are 77 549 stars left. Stars with [Fe/H] $ > -0.7$ dex mainly belong to disk \citep{2013MSAIS2511H, 2014AA562A71B, 2014AA567A5R, 2015MNRAS.453..758H},
and we think it is not suitable for our method to decompose halo from those stars. There are still some stars with [Fe/H] $ < -0.7$ dex belong to the disk, but it is acceptable and we will keep in mind in later analysis. Finally, with metallicity cut, there are 3067 stars left in our sample. GalPot \citep{mcm17} has been used to calculate orbital parameters such as energy \textit{E}, angular momentum \textit{L}, maximum height \textit{Z} of orbit \textit{Zmax}, guiding radius $R_G$ and eccentricity \textit{ecc}. Eccentricities are computed as $ecc = \frac{R_{apo}-R_{per}}{R_{apo}+R_{per}}$ in which $R_{apo}$ and $R_{per}$ are respectively the orbital apocenter and pericenter. The Galaxy potential chosen to calculate these values is the default potential called "PJM17\_best.Tpot" supplied by GalPot.

\begin{figure}
   \centering
   \includegraphics[width=0.8\textwidth, angle=0]{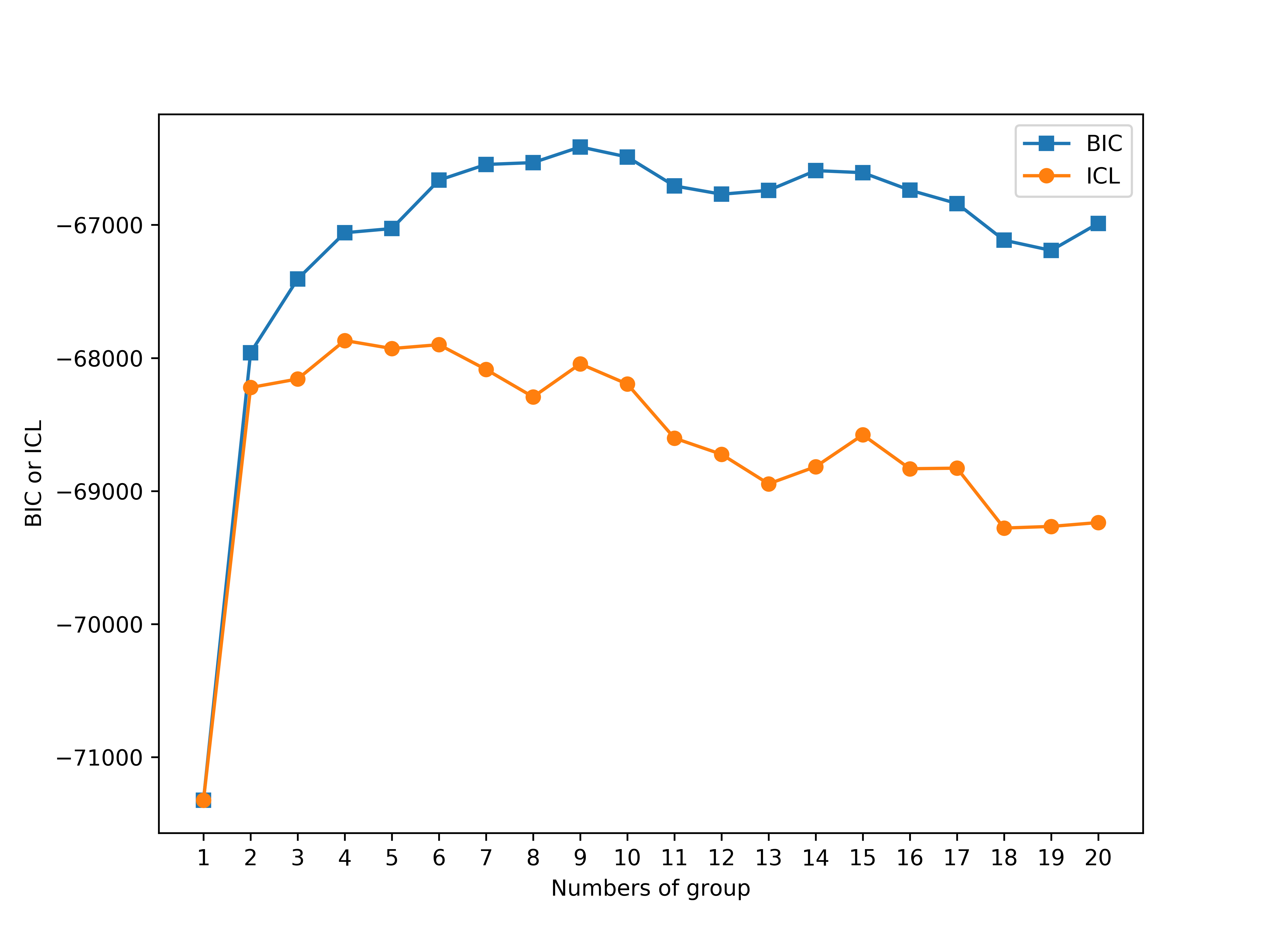}
   \caption{BIC and ICL distribute with numbers of components. }
   \label{fig:bic}
   \end{figure}

R package Mclust \citep{scr16} from R\citet{rct2020} has been used for the model-based clustering, which allows modelling of data as a Gaussian finite mixture with different covariance structures as well as different numbers of mixture components. Each component of a finite mixture density is associated with a group or cluster. The Gaussian mixture model assumes a multivariate distribution for each component and these components are assumed to have ellipsoidal distributions in parameter coordinates \citep{mcl00,fru19}. The number of mixing components and the covariance parameterization are usually selected using the Bayesian Information Criterion (BIC) \citep{sch78,fra98}, which puts its first priority on approximating the density rather than the number of groups. To solve this problem, \citet{bie00} put forward the integrated complete-data likelihood (ICL) criterion, which penalises the BIC through an entropy term by measuring overlapping area to obtain good performance in selecting the number of clusters. When we apply the Gaussian mixture model method to our sample, both BIC criterion and ICL criterion suggest nine mixing components (groups). Figure \ref{fig:bic} shows BIC and ICL values distribute with numbers of components. The BIC criterion suggests nine groups, while it represents a local maximum for ICL criterion. The real maximum of ICL criterion is at four, and the four components are the canonical thin disk, the thick disc, the in-situ born halo and the accreted halo. We choose nine groups because we attempt to study components of the galactic halo rather than main components of the galaxy.

Parameters of Gaussian mixture models are obtained via the EM algorithm \citep{dem77,mcl04}. The EM algorithm is a widely used algorithm which has reliable global convergence under fairly general conditions. However, the likelihood surface in mixture models tends to have multiple modes, and thus initialisation of EM is necessary to produce sensible results when started from reasonable starting values \citep{wu83}. In the R-Mclust package, the EM algorithm is initialised using the partitions obtained from model-based hierarchical agglomerative clustering (MBHAC). It obtains hierarchical clusters by recursively merging the two clusters with smallest decrease in the classification likelihood for Gaussian mixture model \citep{ban93,fraraf98}. The underlying probabilistic model used by MBHAC is shared by both the initialisation step and the model fitting step. When applied to coarse data as ours, the MBHAC approach has a problem that the final EM solution depends on the ordering of the variables. This problem has been solved by using the method presented by \citet{scr15}. Before applying the MBHAC at the initialisation step, the data is projected through a suitable transformation, which enhances separation among groups or clusters. Once a reasonable hierarchical partition is obtained, the EM algorithm is run using the data on the original scale. Via this approach, different orders of variables won't affect fitted model any more and the result becomes stable.

\section{Results}

%\begin{sidewaystable}
\begin{table}
\begin{center}
\caption[]{Parameters of 9 groups.\label{tab:gps}}

%%Please Capitalize the First Letter of Each Notional Word in table's caption

 \begin{tabular}{ccccccccccccc}
  \hline\noalign{\smallskip}
number & [Fe/H] & [Mg/Fe] & $V_\textrm{R}$ & $V_\phi$ & counts & $V_z$ & $Zmax$ & $R_G$ & $ecc$ & $E$ & $L_z$ & comments \\
~ & ~ & ~ & km\,s$^{-1}$ & km\,s$^{-1}$& ~ & km\,s$^{-1}$ & kpc & kpc & ~ & km$^2$\,s$^{-2}$ & kpc km\,s$^{-1}$&  ~ \\
  \hline\noalign{\smallskip}
1 & $-$1.69 & 0.29 &  10.6 &  27.7 &492 & 6.0  & 7.53& 3.12 &0.65 &$-$16898 &217  & \tiny{extremely outer accreted halo}\\
2 & $-$0.80 & 0.25 &  14.4 & 180.1 &247 & 2.15 & 2.13& 7.16 &0.26 &$-$16563 &1684 & \tiny{cross section}\\
3 & $-$1.15 & 0.31 &   1.2 & 104.0 &387 & $-$2.03& 4.09& 3.92 &0.54 &$-$17575 &847  & \tiny{canonical halo}\\
4 & $-$0.86 & 0.31 &  12.4 & 118.6 &428 & $-$0.16& 2.86& 4.25 &0.52 &$-$17649 &934  & \tiny{inner in-situ halo}\\
5 & $-$1.09 & 0.18 & $-$16.9 &  23.9 &607 & 2.33 & 8.20& 2.13 &0.80 &$-$16094 &192  & \tiny{accreted halo}\\
6 & $-$0.75 & 0.31 &  $-$0.1 & 170.8 &680 & $-$4.02& 2.24& 5.86 &0.35 &$-$17086 &1371 & \tiny{the thick disk}\\
7 & $-$1.45 & 0.12 &  24.4 &  85.9 &69  & $-$20.9& 6.85& 6.53 &0.44 &$-$15381 &1189 & \tiny{accreted stars}\\
8 & $-$0.72 & 0.14 &  $-$7.0 & 223.6 &100 & $-$3.39& 1.63&10.04 &0.13 &$-$15154 &2368 & \tiny{the thin disk}\\
9 & $-$1.41 & 0.17 & $-$23.6 & 157.3 &57  &$-$124.3& 9.66& 5.09 &0.38 &$-$15372 &1180 & \tiny{observational effect}\\
  \noalign{\smallskip}\hline
\end{tabular}
\end{center}
\end{table}
%\end{sidewaystable}

We applied the Gaussian mixture models to our data sample in a joint space of [Fe/H], [Mg/Fe], $V_R$ and $V_\phi$ and got nine groups. Table \ref{tab:gps} lists mean values of nine gaussian models fitted by the R-Mclust package. The first column is the number of groups given by R-Mclust package\citep{ker00,mcl87,mcl14}, while the sixth column lists star counts in each group. Table \ref{tab:apt1} in the Appendix lists covariance matrixes of fitted gaussian distribution of each group with variables ordered as ([Fe/H], [Mg/Fe], $V_R$, $V_\phi$). The rest columns of table \ref{tab:gps} list mean values of some kinematical and dynamical parameters after groups were obtained. With more parameters, there would be more smaller substructures with fewer stars in each group. Since we focus on the main components of the halo rather than small substructures, we chose to use these four parameters that can best describe the main components in our sample.

\begin{figure}
\includegraphics[width=1.0\textwidth, angle=0]{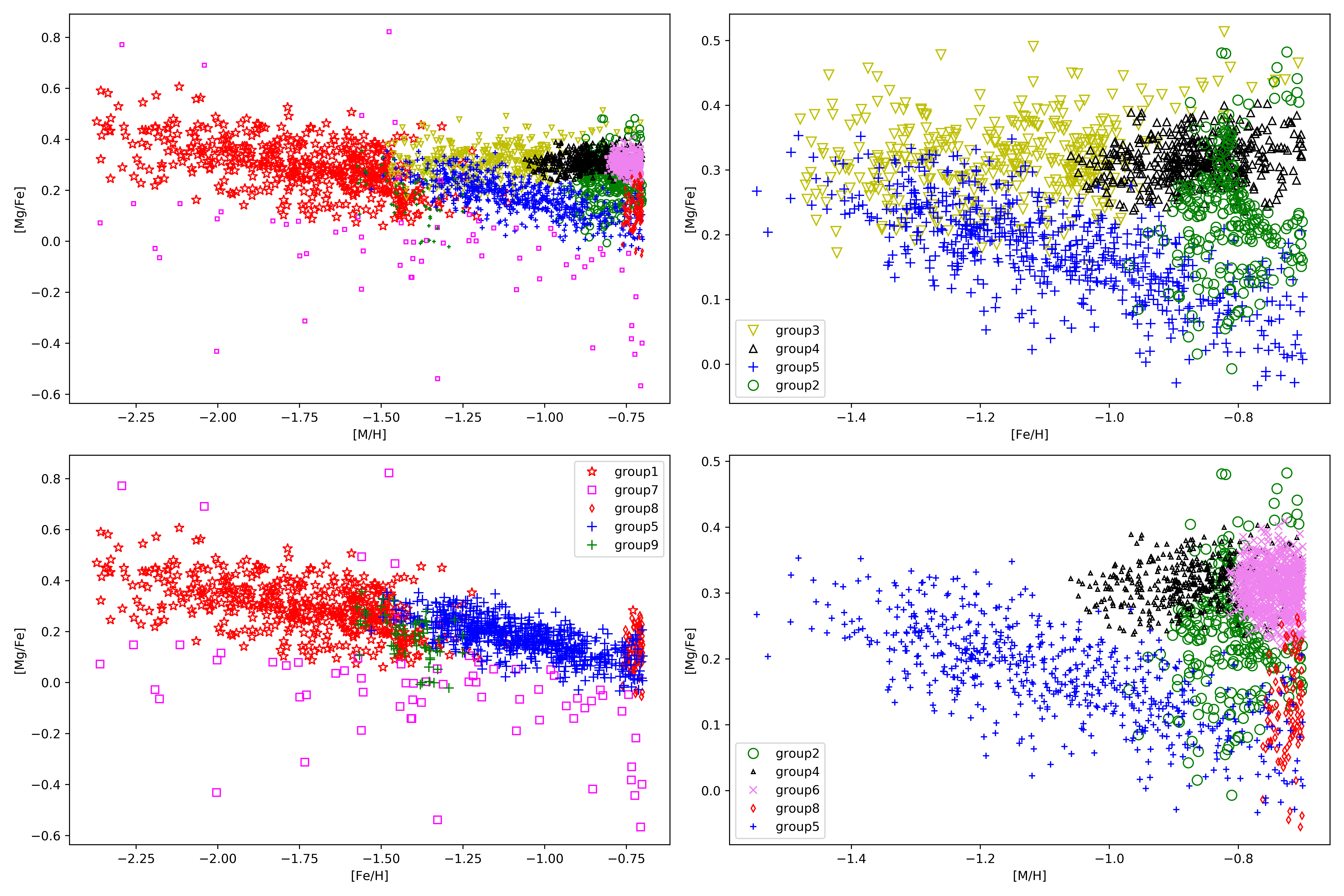}
\caption{The top left panel shows nine groups in the [Mg/Fe] versus [M/H] coordinates, while other three subplots show relatively positions of adjacent groups more clearly.}
\label{fig:mhmg}
\end{figure}

Figure \ref{fig:mhmg} shows nine groups in the [Mg/Fe] versus [Fe/H] coordinate fitted by Gaussian mixture model. The top left panel shows overall distributions of nine groups, while other three panels show relative positions of adjacent groups more clearly. Although Gaussian distribution is not a good model to describe the chemical plane, because stars counts of groups monotonously increase with the increase of metallicity in our sample, it is enough to pick out main components in the sample. Group 3, group 4 and group 6 have similar mean magnesium abundance values around 0.31 and they form the horizontal high-[Mg/Fe] sequence. Group 5, group 8 and group 9 in the lower panel have relatively lower mean magnesium abundance, and Group 7 has even lower [Mg/Fe] ratios. Group 5 takes up the position of canonical accreted halo in the [Mg/Fe] versus [M/H] coordinate \citep{2015MNRAS.453..758H} and Group 9 may be part of the Helmi stream (see Sect. \ref{sec:9}). Group 8 lies at the metal rich end of group 5 and the metal poor end of the thin disk, while group 6 lies at the metal poor end of the thick disk. Group 1 has an increasing [Mg/Fe] with decreasing metallicity and thus we classify it as the extension of the low-[Mg/Fe] sequence.

\subsection{Group 9}\label{sec:9}

\begin{figure}
\includegraphics[width=1.0\textwidth, angle=0]{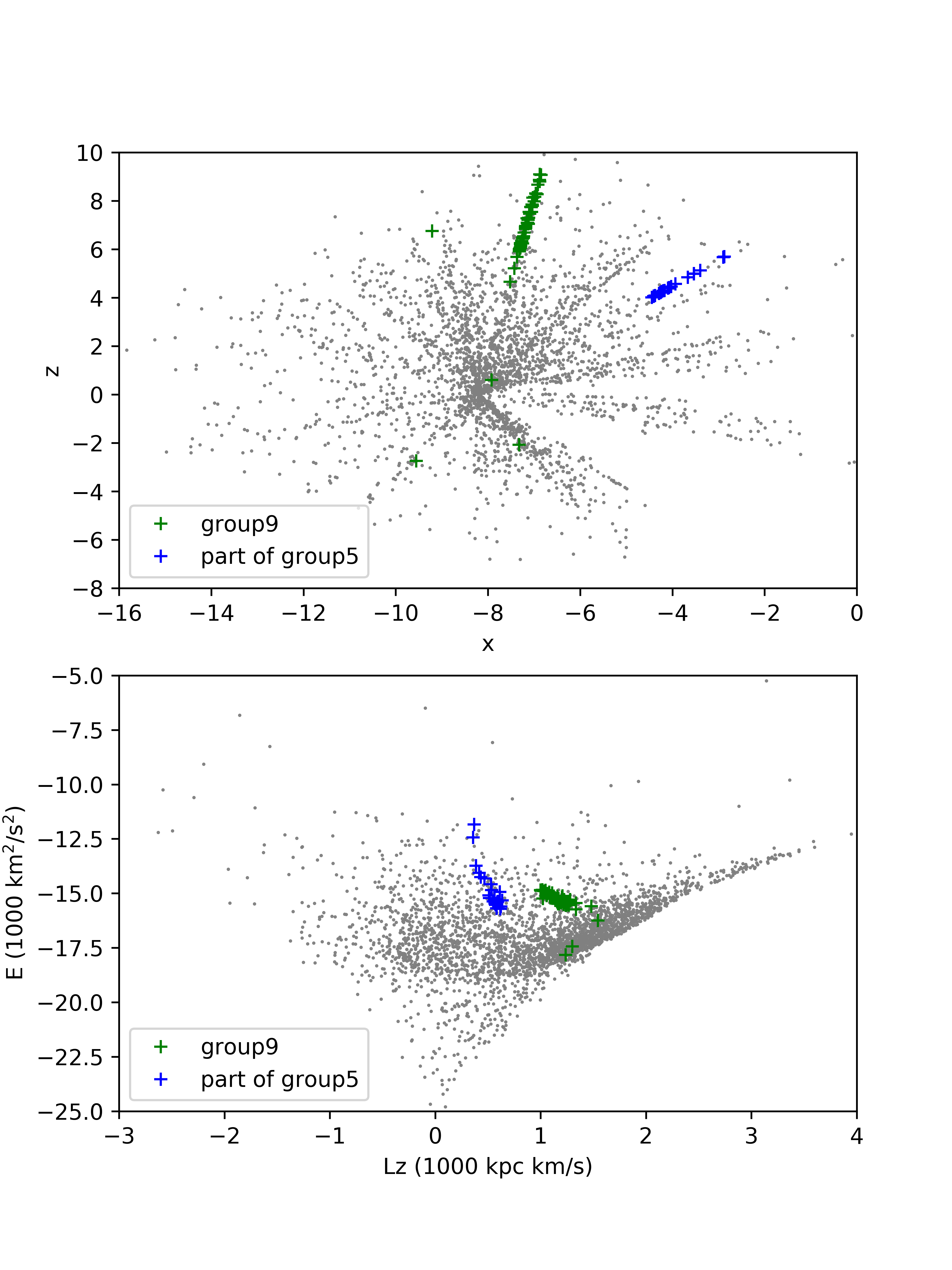}
\caption{Group 9 (green plus) in the (\textit{x, z}) and (\textit{Lz, E}) coordinates. Those blue plus are the small over-density in top left of group 5 in figure \ref{fig:58}. Those background grey dots represent the total sample.}
\label{fig:group9}
\end{figure}

Figure \ref{fig:group9} shows distributions of group 9 (green plus) in the (\textit{x, z}) and (\textit{Lz, E}) coordinates and background grey dots represent the total sample. Grey dots in the upper panel of figure \ref{fig:group9} shows distributions of our total sample in the \textit{x-z} positional space. As shown in the upper panel, group 9 has a almost continuous linear distribution in the \textit{x-z} plane and it is the last group ordered by statistical significance. Its position in the \textit{E} versus \textit{Lz} coordinate is very close to Helmi streams \citep{hel99,kop18,kop19b,2020ApJ...901...48N,2020arXiv200204340H} and we conjecture their observations are related to the Helmi streams. The detection of the Helmi stream is interesting, which may indicate that this method is powerful enough to pick out even small groups with related chemical and kinematic properties. Though we have tried to exclude stars from obvious moving groups which apparently clumped in velocity coordinate when selected the sample, there are still some left in our sample. In a word, we think group 9 is caused by observational effect and do not discuss about it later.

\subsection{Group 3, group 4 and group 6}

\begin{figure}
\includegraphics[width=1.0\textwidth, angle=0]{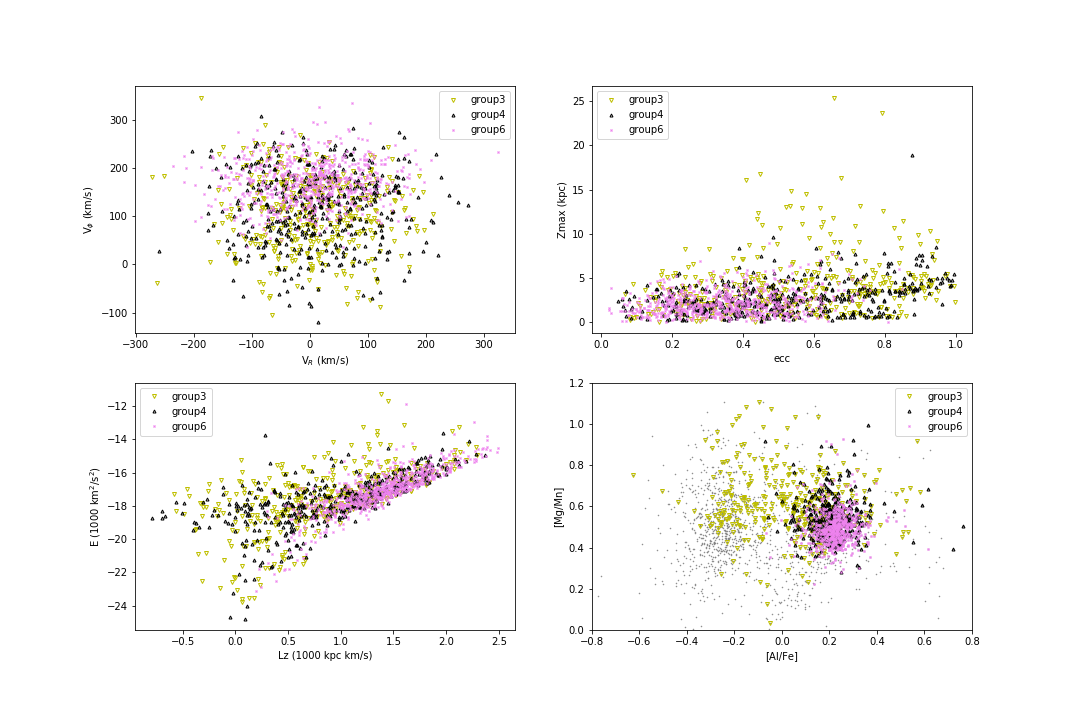}
\caption{Group 3, group 4 and group 6 in ($V_R$, $V_\phi$), (\textit{ecc}, \textit{Zmax}), (\textit{Lz}, \textit{E}) and ([Al/Fe], [Mg/Mn]) coordinates and background grey dots represent our total sample.\label{fig:346}}
\end{figure}

The high-[Mg/Fe] sequence is separated into threes groups (3,4,6) in our sample by the Gaussian mixture model. In the top left panel of figure \ref{fig:346}, group 6 shows thick-disk like kinematics with few retrograding stars. Group 4 and group 3 have very similar distribution ranges and similar scatters in the $V_\phi$ versus $V_R$ coordinate, which means kinematically they can not distinguish from each other. The top right panel of figure \ref{fig:346} shows \textit{Zmax} versus eccentricity distribution of group 3, group 4 and group 6. Group 3 can reach higher region than group 4 and group 6. Most stars of group 6 have roundish orbits and can not run out of disk region. Thus group 6 represents the thick disk while group 3 represents the intermediately outer dual-mode halo or the canonical halo \citep{2015MNRAS.453..758H}. In the bottom left subplot, group 4 have almost similar distribution range as group 3, except its energy is slightly smaller than that of group 3 on the whole. To cleanly distinguish halo component and disk component, the last panel shows those three groups in the [Mg/Mn] versus [Al/Fe] coordinate \citep{2015MNRAS.453..758H}. In the last panel group 4 and group 6 both mainly clumps in the disk region and can not distinguish from each other, While distribution of group 3 reaches to the disk region. Group 4 is tightly related to the disk region that it seems like smoothly extend out of the disk region in all parameter spaces. Since group 4 represents the inner in-situ halo in kinematics but has strong connection to the thick disk in chemistry, we suggest that part of it comes from disk heating \citep{1993ApJ...403...74Q,dim19,2020MNRAS.497.1603G,2020MNRAS.494.1539Y}.

\subsection{Group 5 and group 8}

\begin{figure}
\includegraphics[width=1.0\textwidth, angle=0]{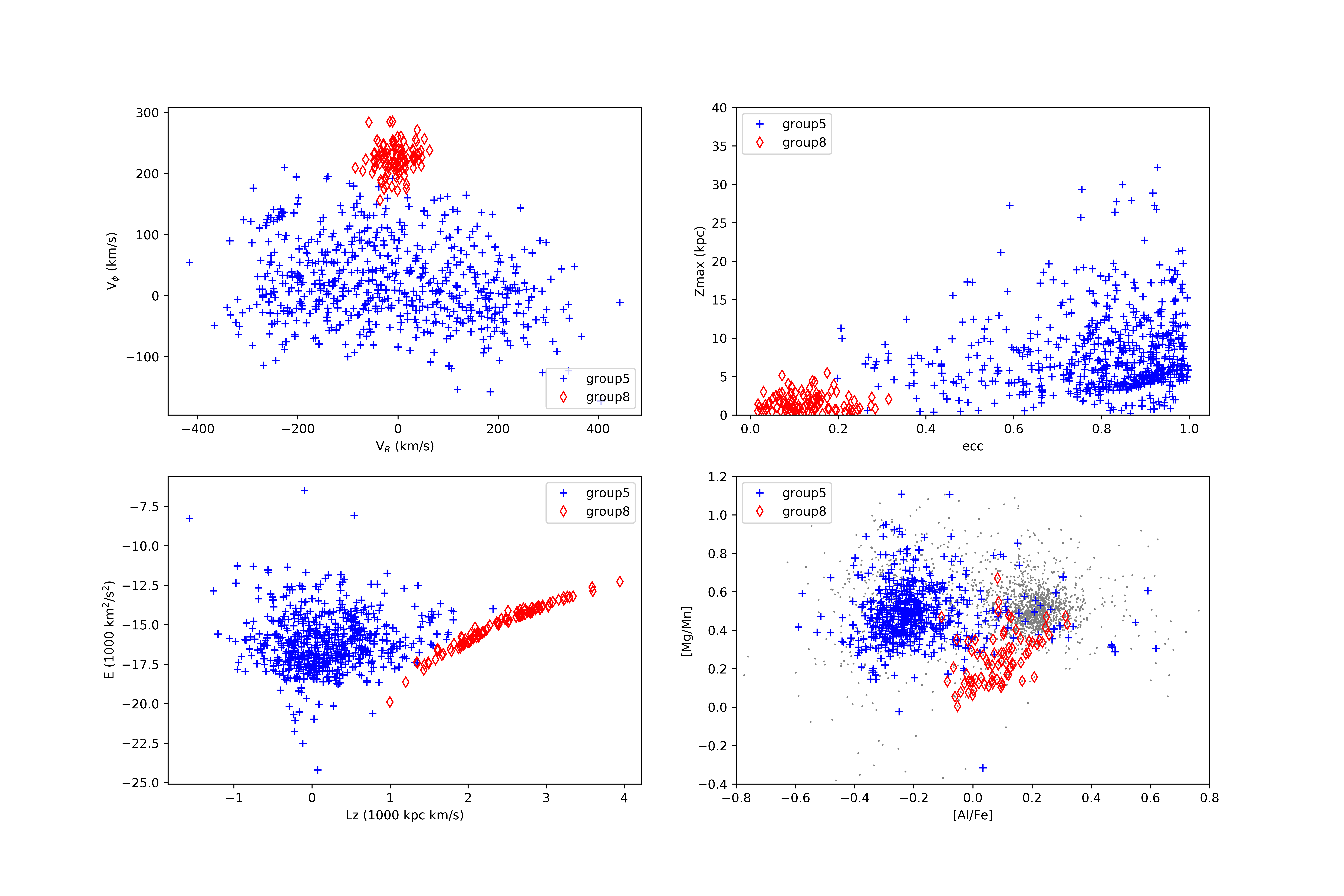}
\caption{Group 5 and group 8 in ($V_R$, $V_\phi$), (\textit{ecc}, \textit{Zmax}), (\textit{Lz}, \textit{E}) and ([Al/Fe], [Mg/Mn]) coordinates and background grey dots represent our total sample.\label{fig:58}}
\end{figure}

In figure \ref{fig:58}, group 8 locates at the metal rich end of group 5 and the metal poor end of the thin disk. However, their distributions in kinematical and dynamical coordinates clearly separate from each other. With high $V_\phi$ speed, low eccentricity and small \textit{Zmax}, group 8 represents the thin disk component in our sample. Group 5 represents the canonical accreted halo component \citep{2015MNRAS.453..758H} according to its position in figure \ref{fig:mhmg}. Group 5 is mainly composed of Gaia-Enceladus-Sausage, but includes more than just it. Since our sample stars are within $z \approx $10 kpc, we conformed that Gaia-Enceladus-Sausage is the dominant component of the halo \citep{2020ApJ...901...48N} in this spatial range. There is a megascopic group of stars clumped in the top left region of group 5 in the $V_\phi$ versus $V_R$ coordinate. As is shown in figure \ref{fig:group9}, They have a continuous distribution in the \textit{x-z} plane like group 9 \citet{zhao21} and it may be caused by observational effect too. Chemical abundances of these stars is very close to group 9 (the Helmi stream) too. But they have different $V_R$ coordinates and larger eccentricity distributions than stars in group 9. We think they may be related to the Helmi stream and it is normal that some stars from the Helmi stream or other streams are classified into the accreted halo (group 5). Group 5 not only cleanly separates from the thin disk in kinematical and dynamical coordinates, but also has different distribution from the thin disk in the [Mg/Mn] versus [Al/Fe] coordinate. Unlike group 3 in figure \ref{fig:346}, chemical distribution of group 5 is apparently separated from the disk region with only few stars scattering in the disk region. Comparing our bottom left panel with selection criteria in \citet{2020ApJ...901...48N}, there are many dynamical substructures in the accreted halo such as Gaia-Enceladus-Sausage, Wukong, Helmi streams and Aleph.

\subsection{Group 1 and group 7}

\begin{figure}
\includegraphics[width=1.0\textwidth, angle=0]{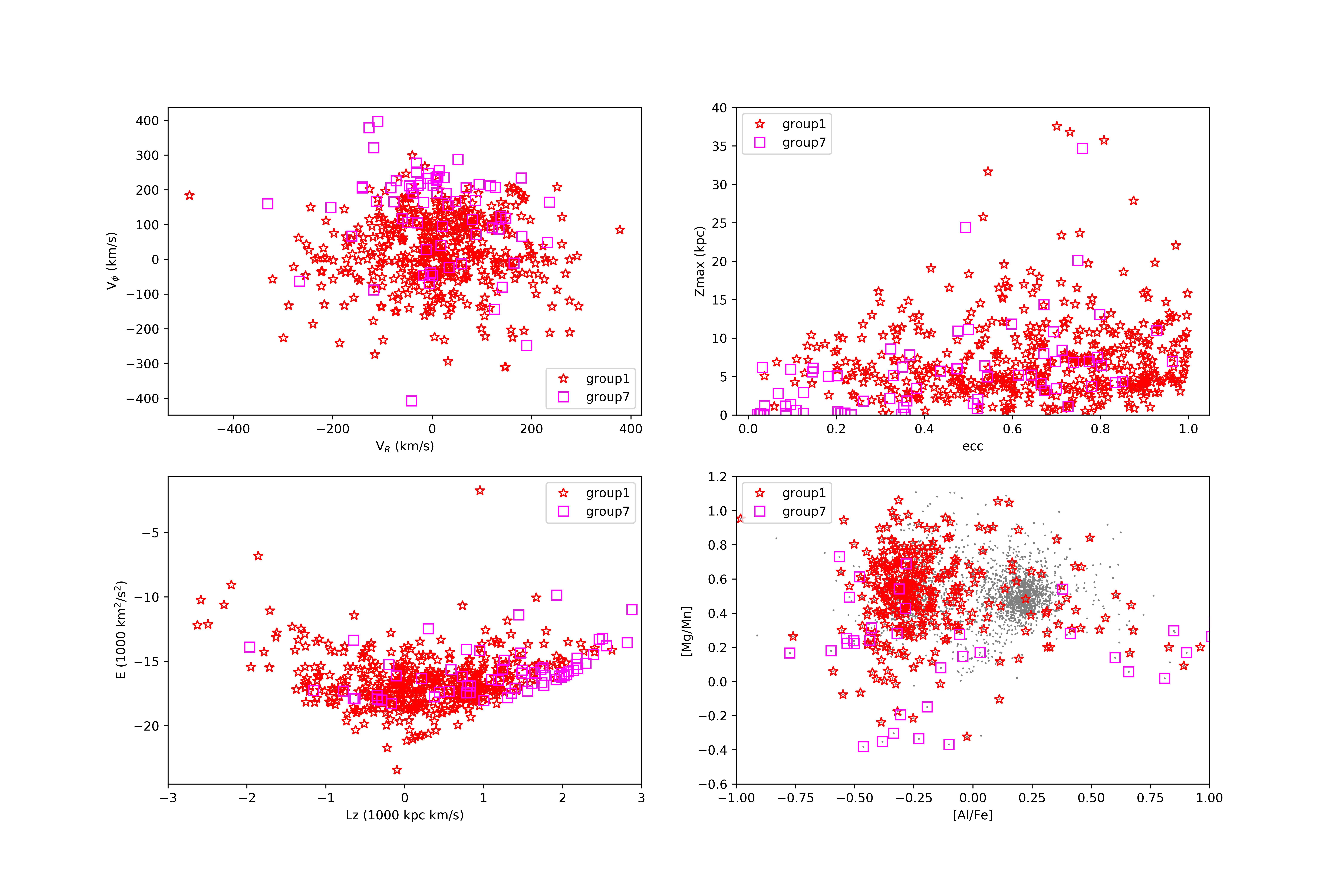}
\caption{Group 1 and group 7  in ($V_R$, $V_\phi$), (\textit{ecc}, \textit{Zmax}), (\textit{Lz}, \textit{E}) and ([Al/Fe], [Mg/Mn]) coordinates and background grey dots represent our total sample.\label{fig:17}}
\end{figure}

Figure \ref{fig:17} shows group 1 and group 7 in kinematic and dynamical coordinates. Group 1 represents the metal poorest part of the halo while group 7 is composed of stars with poorest magnesium abundances in our sample. They both have large ranges of velocity distributions and spatial distributions. In the \textit{E} versus \textit{Lz} coordinate, group 1 overlaps with parts of some dynamical substructures such as Gaia-Enceladus-Sausage, Wukong, Arjuna, Sequoia, I'itoi, Thamnos and Aleph \citep{2020ApJ...901...48N}. Group 1 has large scatters for both kinematics and chemical abundances distributions. And in all those subplots of figure \ref{fig:17}, group 1 shows smooth distributions. Within group 1, stars born in the Milky Way can not be distinguished from those accreted from other galaxies either chemically or kinematically. The bottom right panel of figure \ref{fig:17} shows [Mg/Mn] versus [Al/Fe] distributions of group 1 and group 7 and those background grey dots represent our total sample. Main parts of these two groups clearly distribute away from the disk region and group 7 seems distribute slightly father away from the disk than group 1. Group 7 has even poorer metallicity and lower Magnesium abundance than group 5, the accreted halo. Based on these properties, we suggest that group 1 represents the extremely outer accreted halo, while group 7 might be related to dwarf galaxies due to lower [Mg/Fe] than the accreted halo, typical for dwarf galaxies.

\subsection{Group 2}

\begin{figure}
\includegraphics[width=1.0\textwidth, angle=0]{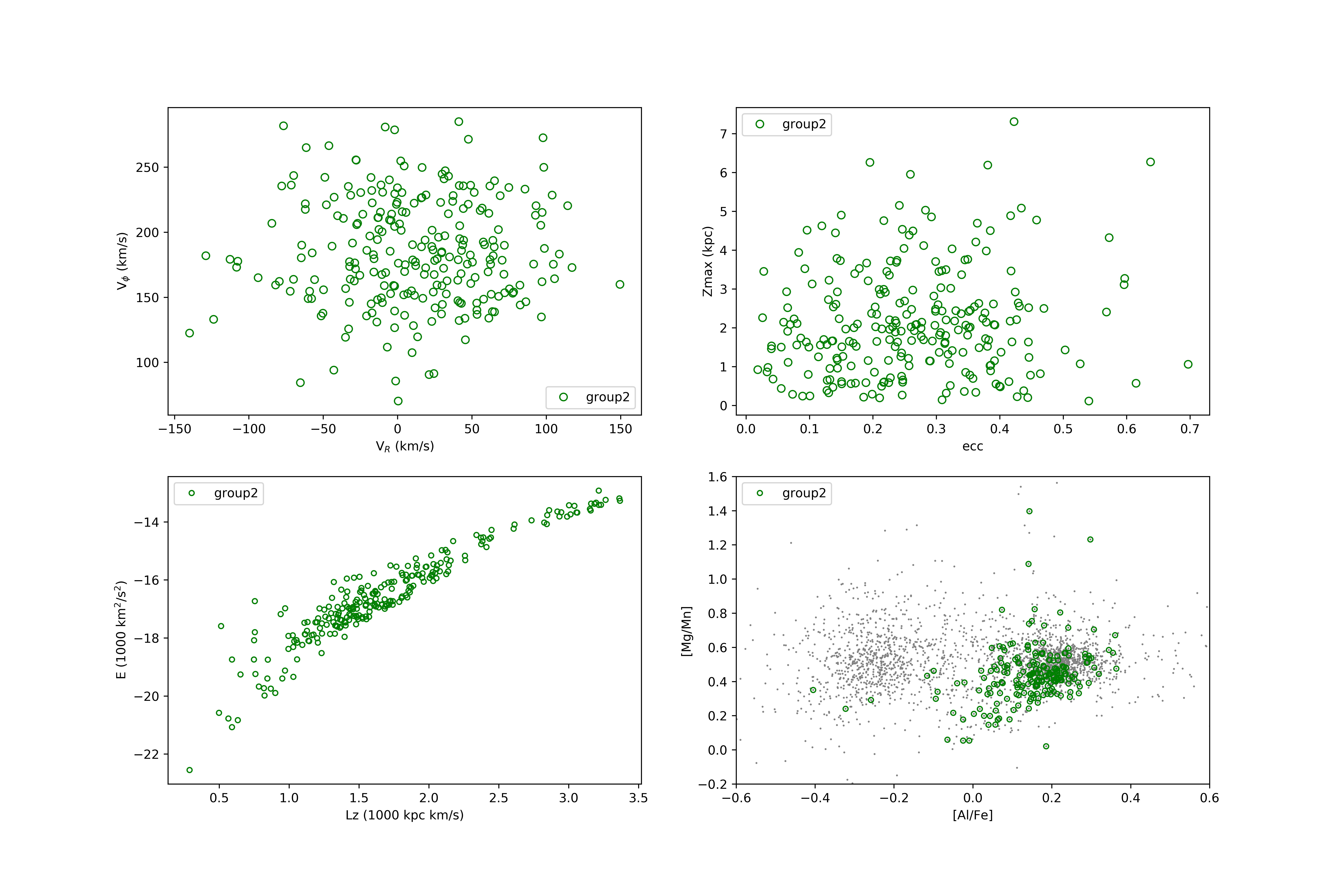}
\caption{Group 2 in ($V_R$, $V_\phi$), (\textit{ecc}, \textit{Zmax}), (\textit{Lz}, \textit{E}) and ([Al/Fe], [Mg/Mn]) coordinates and background grey dots represent our total sample.\label{fig:2}}
\end{figure}

In figure \ref{fig:mhmg}, group 2 distributes over both high-[Mg/Fe] branch and low-[Mg/Fe] branch and it connects the thin disk, the thick disk, the in-situ halo and the accreted halo. The kinematic and dynamical distributions of group 2 in figure \ref{fig:2} behave like a disk component. Most stars of group 2 have not very large eccentricity and highest distances away from the Galactic middle plane and thus they are mainly born in the disk region. In the last panel, [Mg/Mn] versus [Al/Fe] distribution of group 2 mainly locates at disk region with few low [Al/Fe] or high [Mg/Mn] stars permeating into halo region. As compared with group 8, group 2 has the main component of the thick disk with higher [Al/Fe] or high [Mg/Mn] ratios but it has a tail extending to lower values, overlapping with the thin-disk region. Thus, group 2 represents the cross section of thin disk and the thick disk and may contains some stars from accreted halo. Usually the cross section is not very obvious, but most stars of group 2 distribute close to the Sun in spatial coordinates. The observational effect may have played a role in making it an detectable component in our sample.

\subsection{Discussion}

\begin{figure}
\includegraphics[width=1.0\textwidth, angle=0]{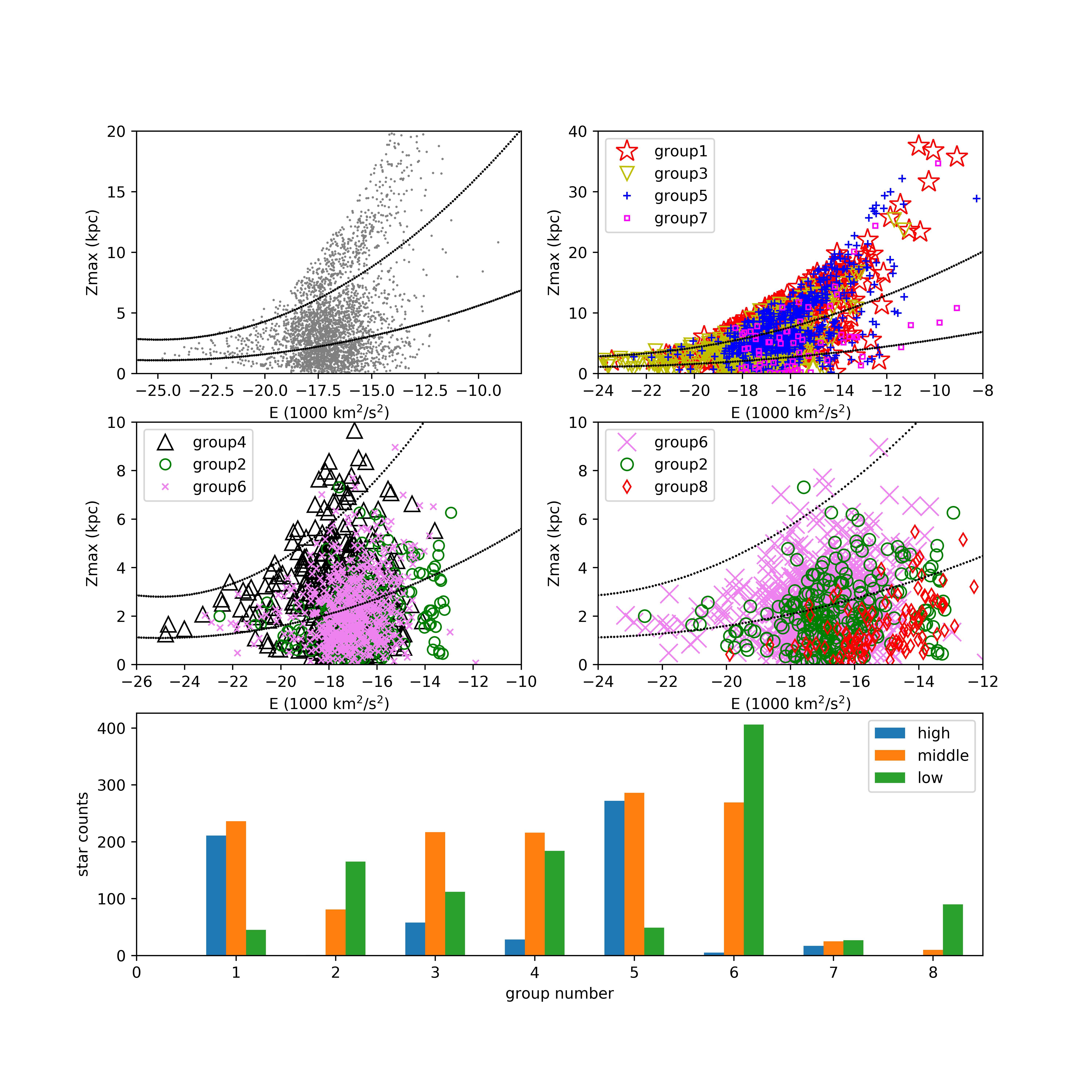}
\caption{\textit{E} versus \textit{Zmax} distributions.\label{fig:ez}}
\end{figure}

Figure \ref{fig:ez} shows those eight groups in \textit{Zmax} versus \textit{E} coordinates. The top left subplot shows the distribution of our total sample, and it apparently distributes into three blocks. Two black dotted curves are subjectively drawn to show boundaries between those three blocks caused by transitions from one orbital family to another \citep{2015MNRAS.451..705M,hay18,2020MNRAS.492.3816A}. The locations of these gaps marked by black dotted curves are determined by the adopted potential and cannot be used to learn about the physical origins of the stars \citep{2015MNRAS.451..705M,2020MNRAS.492.3816A}. Nonetheless, it is still interesting to see some disk stars being scattered to high orbits by resonant effects. The bottom subplot of figure \ref{fig:ez} shows stars orbital distribution of each group in high, middle and low orbital families. We notice that the accreted halo groups, group 1 and group 5 has similar distributions with comparable fraction between the high and middle sections and little contribution from the low section. Group 2 and group 6 have main contribution from the low section but significant fraction from the middle section, which is consistent with the thick disk population. Group 4 shows a larger fraction of the middle section than the low section, similar as the in-situ halo of group 3, and thus indicates an evidence for disk heating. The star number in group 7 is too small to draw conclusion. Group 8 has main distribution in the low section in consistent with the thin disk population.

\begin{figure}
\includegraphics[width=1.0\textwidth, angle=0]{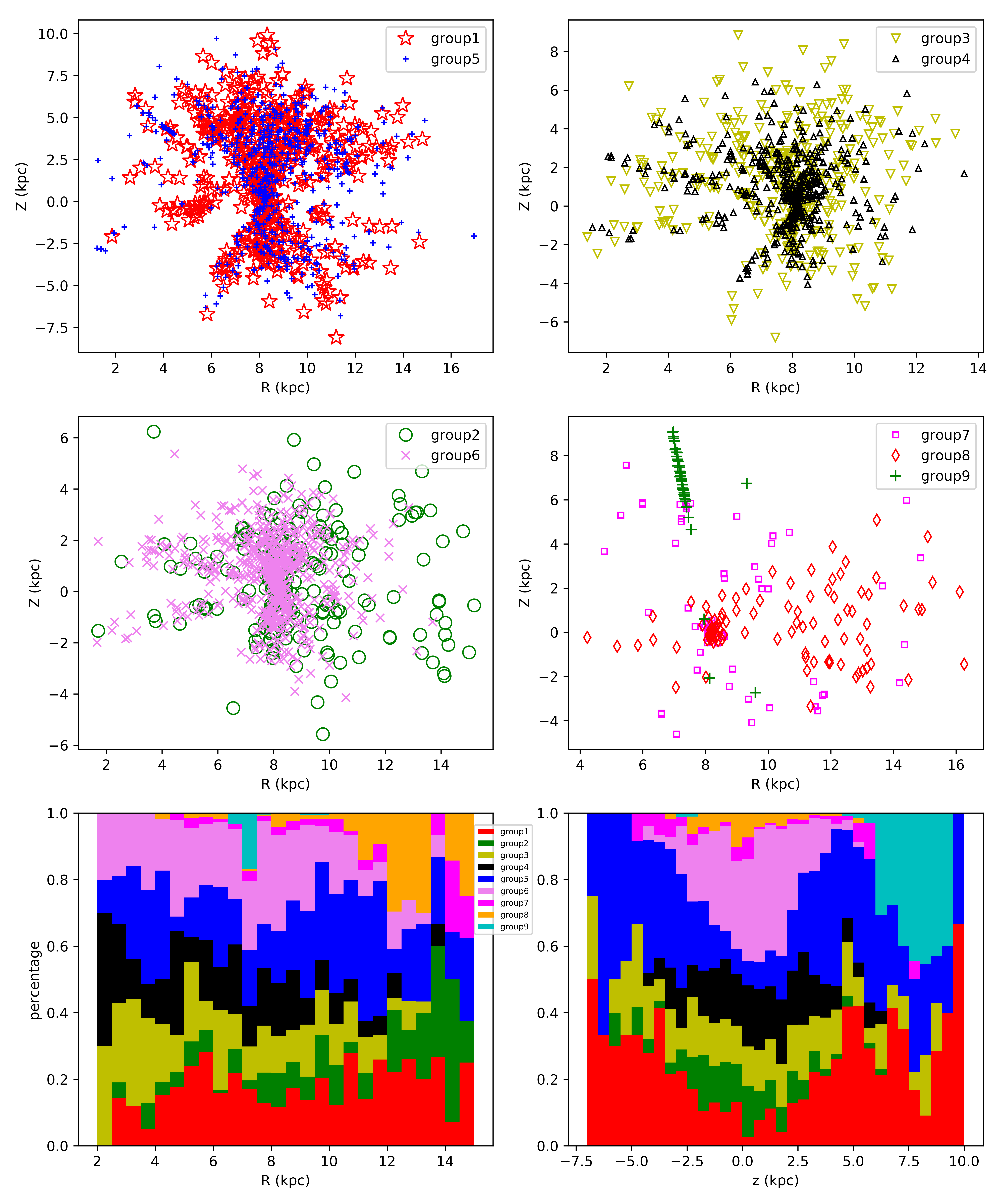}
\caption{Spatial distributions of each group. Bins of 0.5 kpc have been taken along $R$ and $z$ direction to obtain star counts in the bottom two panels. \label{fig:space}}
\end{figure}

Figure \ref{fig:space} shows spatial distributions of our nine groups in $(R, z)$ coordinates. While bottom two stacked bar charts show distributions of percentages each group take up in small bins taken along \textit{$R$} and $z$ direction with size equal to 0.5 kpc. Group 1 and group 5 have the largest spatial distributions while group 6 and group 8 have the smallest spatial distributions, which is consistent with their origins deduced by chemical and kinematical parameters. The spatial distribution of our sample is restricted by observational effect that most stars distribute around the sun as shown in top four panels. But percentage distributions reveal some tendencies. In the bottom left panel, the large $R$ end is mainly composed of group 8 (the thin disk), group 7 (accreted stars), group 5 (accreted halo), group 1 (extremely outer accreted halo) and group 2 (cross section of disks). For group 8, only the bin at 8 kpc contains more than 10 stars, but it still needs explanation why group 8 takes up about twenty percent of our total sample at large $R$ region. As is shown in the bottom left panel, group 4 (inner in-situ halo) take up very small percentages in bins around the large $R$ end relative to other regions, while group 2 and group 8 take up relatively larger percentages at this region. As is known, there is radial abundance gradients of the disk \citep{Shaver1983, lia19}  and at large radius, the flare and warp of the disk make it more easily been observed. Considering our sample is selected by metallicity cut, it is reasonable the thin disk component takes up more percentage at large $R$ region. It is also possible there are several thick disk stars or accreted stars or accreted halo stars mixed in group 8. For radius larger than the Galactic center regions (around 3 kpc), group 1 takes up similar percentages at different Galactic radius and so does group 5. Group 7 takes up larger percentages at larger Galactic radius and we think group 8 might contain several stars from dwarf galaxies too. In the last panel, group 6 (the thick disk component) dominates the Galactic mid-plane region while group 1 and group 5 dominate regions far away from the Galactic mid-plane. Group 4 distributes around the Galactic mid-plane while group 3 (canonical halo) does not show clear preference for $z$ distribution. The inner in-situ halo (group 4) takes up similar spatial range in $z$ direction as disk components (group 6, group 2 and group 8).

\section{Conclusion}

A sample of stars with reliable stellar parameters in the ranges of [Fe/H] $ < -0.7$ dex, $-0.5$ $< $ [Mg/Fe] $ < 0.6$ dex, $Zmax < 90$ kpc, $E < 0$ $km^2\,s^{-2}$ and $\mid Lz \mid  < 5000$ kpc km\,s$^{-1}$ is selected from the APOGEE DR16 catalogue. Gaussian mixture model has been applied to this sample by using both chemical abundances parameters of [Mg/Fe] and [M/H] and two velocity components of \textit{$V_R$, $V_\phi$}. Nine groups are detected, in which group 9 is part of the Helmi stream caused by observational effect. We analysed distributions of each group on the [M/H] versus [Mg/Fe], $V_R$ versus $V_\phi$, the \textit{Zmax} versus \textit{eccentricity}, \textit{Energy} versus \textit{Lz} and [Mg/Mn] versus [Al/Fe] diagrams in order to trace their origins. Group 1 represent the extremely outer accreted halo, while group 7 comes from dwarf galaxies. Comparing the position of group 1 in the \textit{Lz} versus \textit{E} diagram with the streams of \cite{kop19} indicate that group 1 includes many accreted stars from Sequoia, Thamno 1 and Thamno 2 substructures as well as many local born halo stars. Group 5 is the accreted halo dominated by the well known Gaia Enceladus/Sausage, which is clearly separated from the in-situ halo of group 3 in both chemical ([Mg/Fe] and [Al/Fe]) and kinematic spaces ($V_R$ versus $V_\phi$ and the \textit{Lz} versus \textit{E} diagrams). We may see a signature of disk heating in group 4, part of which has a halo-kinematics in the $V_R$ versus $V_\phi$ diagram but disk-chemistry in the [Mg/Mn] versus [Al/Fe] diagram. Group 6 and group 8 show typical properties for the classical thick disk and the thin disk respectively, while group 2 may consists mainly thick disk stars with a small fraction of thin disk stars. In the \textit{E} versus \textit{Zmax} diagram, three sections are found and the relative fractions provide further support on the similar origin of group 1 and group 5 from the accreted halo, of group 2 and group 6 from the thick disk, as well as the signature of disk heating for group 4 (as compared with the in-situ halo of group 3). These results indicate that the Galactic halo has complicated assembly history and it not only interacts but also strongly mixed with other components of the Galaxy and satellite dwarf galaxies. In addition, dynamical evolution and chemical evolution of the Galaxy are entangled, and these factors should be considered together in the simulation works in the future.

\begin{acknowledgements}
We thank the reviewer for his/her constructive suggestions. This study is supported by the National Natural Science Foundation of China under grants No. 11988101, 11625313, 11973048, 11927804, 11890694 and National Key R\&D Program of China No. 2019YFA0405502 and the Fundamental Research Funds for the Central Universities under grants No. 292020001734 (E0E48956).

Funding for the Sloan Digital Sky Survey IV has been provided by the Alfred P. Sloan Foundation, the U.S. Department of Energy Office of Science, and the Participating Institutions. SDSS-IV acknowledges support and resources from the Center for High-Performance Computing at the University of Utah. The SDSS web site is www.sdss.org. SDSS-IV is managed by the Astrophysical Research Consortium for the Participating Institutions of the SDSS Collaboration including the Brazilian Participation Group, the Carnegie Institution for Science, Carnegie Mellon University, the Chilean Participation Group, the French Participation Group, Harvard-Smithsonian Center for Astrophysics, Instituto de Astrof\'isica de Canarias, The Johns Hopkins University, Kavli Institute for the Physics and Mathematics of the Universe (IPMU) / University of Tokyo, the Korean Participation Group, Lawrence Berkeley National Laboratory, Leibniz Institut f\"ur Astrophysik Potsdam (AIP), Max-Planck-Institut f\"ur Astronomie (MPIA Heidelberg), Max-Planck-Institut f\"ur Astrophysik (MPA Garching), Max-Planck-Institut f\"ur Extraterrestrische Physik (MPE), National Astronomical Observatories of China, New Mexico State University, New York University, University of Notre Dame, Observat\'ario Nacional / MCTI, The Ohio State University, Pennsylvania State University, Shanghai Astronomical Observatory, United Kingdom Participation Group, Universidad Nacional Aut\'onoma de M\'exico, University of Arizona, University of Colorado Boulder, University of Oxford, University of Portsmouth, University of Utah, University of Virginia, University of Washington, University of Wisconsin, Vanderbilt University, and Yale University.

This work has made use of data from the European Space Agency (ESA) mission Gaia (https://www.cosmos.esa.int/gaia), processed by the Gaia Data Processing and Analysis
Consortium (DPAC, https://www.cosmos.esa.int/web/gaia/dpac/consortium). Funding for the DPAC has been provided by national institutions, in particular, the institutions participating in the Gaia Multilateral Agreement.

\end{acknowledgements}

\appendix
\section{Gaussian finite mixture model}
Let $x = ( x_1, x_2, ... , x_i , ... , x_n ) $ be a sample of n independent identically distributed observations from a probability density function through a finite mixture model of g Gaussian components, which takes the following form

\begin{equation*}
   f(x_i, \Psi)= \sum_{i=1}^{g} \pi_i \phi(x_i; \mu_i, \Sigma_i)
\end{equation*}

Where $ \Psi = (\pi_1, ... , \pi_g-1, \xi^T)^T$ are parameters of the mixture model. $\xi$ contains the elements of the component mean $\mu_i$ and covariance matrices $\Sigma_i$, while $\pi_i$ is the mixing weight or probability of each component. For our case, each observation is composed of four variables ordered as ([Fe/H], [Mg/Fe], $V_R$, $V_\phi$). Mean values of nine multivariate Gaussian distributions have been listed in table \ref{tab:gps}, while table \ref{tab:apt1} lists covariance matrixes of these nine components.

\begin{table}
\caption[]{Covariance matrix of fitted gaussian distributions. They are ordered by group numbers with four used variables for each covariance matrix ordered as ([Fe/H], [Mg/Fe], $V_R$, $V_\phi$).}
\label{tab:apt1}

\begin{eqnarray}
\left(
\begin{array}{rrrr}
~0.09349142 & ~~~~$-$0.01670629~~&~ ~   $-$0.0510382~  & ~ ~$-$5.4292427\\
 $-$0.01670629 &~~~ 0.01039478 ~~& ~   0.3099455 ~&~    0.7175462\\
 $-$0.05103820 &~~~ 0.30994546 ~~&~      14693.3~ &~  288.5084\\
 $-$5.42924271 &~~~ 0.71754620~~& ~  288.508 ~&~ 10116.28
\end{array}\right)  \\
\left(
\begin{array}{rrrr}
0.0034185456~ &~ 0.0004206848~~~& ~ ~  0.3631803~ &~  ~$-$0.7387926\\
0.0004206848~ & 0.0078440981~~~ &   1.2687077 ~ &~ $-$1.2341388\\
0.3631803393 ~& 1.2687077~~~& 3059.0287~ &~  23.3533386\\
$-$0.7387926~& $-$1.23413881~~~ &  23.3533386~ &~1826.7034
\end{array}
\right) \\
\left(
\begin{array}{rrrr}
0.03838573~ &~~ 0.003348854 &~~ ~~ $-$5.2917447~~ & ~  ~3.4282508\\
 0.00334885~ &~ 0.004225013 & ~~~ $-$0.2554129 ~~&~  $-$0.4454307\\
$-$5.29174471~ &~$-$0.255413 &~~~7421.4537~~&~ $-$587.3283\\
 3.42825076~&~ $-$0.445430748 &~~~$-$587.3283~~ &~6411.704
\end{array}
\right)\\
\left(
\begin{array}{rrrr}
0.0075656207 &~ 0.0005721973 ~ &~~~ ~$-$1.0612834  &~~ $-$2.9071771\\
0.0005722 & 0.0011567533 ~ &  ~~~0.1007521  &~ $-$0.2963813\\
$-$1.0612833 & 0.1007521000~ &~~~7554.22 &~ 221.595\\
 $-$2.907177 &$-$0.29638133~ &~~~ 221.595&~ 5618.139
\end{array}
\right) \\
\left(
\begin{array}{rrrr}
 0.04021166~ ~& ~$-$0.01154754~~~~ &~    ~1.2517914& ~ ~$-$0.4534044\\
$-$0.01154754 ~~& 0.00621082~~~~ & ~   0.4413265 & ~   $-$0.7344783\\
1.25179140~ ~& 0.4413265~ ~~~ & ~  27383.1653 &~ $-$1998.32\\
$-$0.45340438 ~~& $-$0.7344783~~~~ & ~$-$1998.32    &~4412.365
\end{array}
\right) \\
\left(
\begin{array}{rrrr}
0.00093047~~ &~$-$0.00002497~~~~ &~0.01589945 &~~$-$0.0076799\\
$-$0.00002497~~&  0.001245~~~~ &0.02610716 &~ 0.08878\\
0.015899 ~~& 0.02610716 ~~~~& 6691.895  &~313.2556\\
$-$0.0076798~~& 0.08878387~~~~ &313.2556 &~ 2258.296
\end{array}
\right)  \\
\left(
\begin{array}{rrrr}
0.18996237  ~&~$-$0.05379311~~~~~~ &  ~  1.236969~ & ~  20.72871\\
$-$0.05379311~ &  0.06226352~~~~~~&     5.277472~&   $-$11.82159\\
1.23696882 ~ & 5.27747232~~~~~~ &10628.428~& $-$3281.21520\\
20.72870640 ~&$-$11.82159321 ~~~~~~&$-$3281.2~ &18444.13
\end{array}
\right)\\
\left(
\begin{array}{rrrr}
 0.000314823 & ~0.0001351775&  ~-0.1034287 &  ~0.01897870\\
 0.000135178&  0.0059722326&  $-$0.7496104&  $-$0.06235303\\
$-$0.10342873 &$-$0.749610434& 991.1423 &106.8586878\\
 0.0189787& $-$0.062353027 &106.85869 &643.55376
\end{array}
\right) \\
\left(
\begin{array}{rrrr}
0.006454853 ~~&~~$-$0.00370338~~~ &~ 0.01633454 &~~~$-$0.0272713\\
$-$0.00370338 ~~&~ 0.007584~~~& $-$0.0640229&~~  0.01518699\\
0.016334538~~ &~$-$0.0640229 ~~~&52.2878872&~~ 44.091147\\
$-$0.02727136 ~~&~ 0.015187~~~& 44.0911476 &~~73.628824
\end{array}
\right)
\end{eqnarray}

\end{table}

\end{document}